# Metal Doping in Topological Insulators- A Key for Tunable Generation of Terahertz

Prince Sharma[1,2], M.M Sharma[1,2], Mahesh Kumar[1,2], V.P.S Awana[1,2*]

[1]*CSIR-National Physical Laboratory, Dr. K.S. Krishnan Marg, New Delhi-110012, India*
[2]*Academy of Scientific and Innovative Research (AcSIR), Ghaziabad, 201002, India*

## Abstract

The unique surface edge states make topological insulators a primary focus among different applications. In this article, we synthesized a large single crystal of Niobium(Nb)-doped $Bi_2Se_3$ topological insulator (TI) with a formula $Nb_{0.25}Bi_2Se_3$. The single crystal has characterized by using various techniques such as Powder X-ray Diffractometer (PXRD), DC magnetization measurements, Raman, and Ultrafast transient absorption spectroscopy (TRUS). There are (00l) reflections in the PXRD, and Superconductivity ingrown crystal is evident from clearly visible diamagnetic transition at 2.5K in both FC and ZFC measurements. The Raman spectroscopy is used to find the different vibrational modes in the sample. Further, the sample is excited by a pump of 1.90 eV, and a kinetic decay profile at 1.38 eV is considered for terahertz analysis. The differential decay profile has different vibrations, and these oscillations have analyzed in terms of terahertz. This article not only provides evidence of terahertz generation in Nb-doped sample along with undoped sample but also show that the dopant atom changes the dynamics of charge carriers and thereby the shift in the Terahertz frequency response. In conclusion, a suitable dopant can be used as a processor for the tunability of terahertz frequency in TI.

Key Words: Topological Insulators; Self-Flux solid-state reaction route; PXRD and Raman Spectroscopy; Terahertz Tunability; Ultrafast transient absorption spectroscopy.

[*]**Corresponding Author**

Dr. V. P. S. Awana:  E-mail: awana@nplindia.org
Ph. +91-11-45609357, Fax-+91-11-45609310
Homepage: awanavps.webs.com

## Introduction

The usage of terahertz frequency has been increasing due to its incredible applications in the field of security scanning, medical imaging, communication system, information processing, and the scrutiny of pharmacological complexes in the past few years[1–5]. The 0.1-10 THz range is the key to these applications, which has led to the development of robust materials and compact THz instrumentation. The trivial method for generating terahertz is through the high intense femtosecond laser and with the help of target materials[3–6]. Recently, the focus is turned towards the topological insulators as the implausible candidate for target materials[6–12]. The optical, magnetic, and electrical characteristics of TI show the potential for terahertz generation[13–16]. TIs are nematic phase materials characterized by no gap states exhibiting polarized spins at the surface and degenerate spin with full-gap states in bulk[17–20]. Among all reported TIs, Bi-based TIs are the most popular TIs, and it is well established now that doping of impurity elements in these TIs gives rise to the various important phenomenon. TIs doped with magnetic impurities such as Mn, Cr, and Fe have proved to be an excellent platform to observe the anomalous Quantized Hall effect and ferromagnetism[21–24]. Interestingly, the presence of the Van der Waals gap in these materials makes these materials a good choice for element intercalation. Intercalation of elements such as Nb, Cu, Sr, Pd, and Tl have found to dope more carriers in Bi-based TIs and leads to the occurrence of Superconductivity [21,25–31]. The effect of doping on electrical and magnetic properties of TIs is studied significantly, but its effect on the optical properties of TIs has not studied to that extent.

Nb-doped crystal shows the highest superconducting critical temperature, among others. The 0.25 doping spectacles the incredible results as spontaneously breaks the time-reversal symmetry due to the formation of Majorana bound states[28,31–33]. The states are non-Abelian particles and have a potential of topologically protected qubits in quantum computation. The Nb-doped is essential because of its non-magnetic nature as well the same intercalates between the Van-der wall layers easily. Apart from its superconducting and electrical properties, the focus has turned towards the optical and terahertz enchantment of Bi-based TIs. The investigation on Cu doped $Bi_2Se_3$ presented phonon frequency shift[20,22,29,30,34]. The redshift confirmed the intercalation of the Cu atom between every pair of quintuple layers. The effect of dopants in the terahertz generation has become extremely interesting. In this article, Nb-doping in $Bi_2Se_3$ TI has been investigated, particularly the structural changes, and thereby, Nb induced tunable character of changing the terahertz frequency is studied using the Raman and the transient absorption ultrafast spectroscopy. This article substantiates the generation of a broad range of terahertz frequencies with the help of a metal dopant in a TI crystal.

**Experimental Details**

The solid-state reaction route synthesizes the Nb-doped TI single crystal. The self-flux method has been used in which the atoms above melting points self-induce the reaction. The optimal stoichiometric ratio of Nb, Bi, and Se are taken and grind using the mortar pester in the mbraun glove box at argon atmosphere. The sample is then pelletized in the quartz tube and placed in the box furnace at optimized heat treatment, as shown in our previous paper[33]. The synthesized single crystal has been characterized by different techniques such as PXRD and Raman spectroscopy. DC magnetization measurements have been made by using Quantum Design PPMS under Zero Field Cooled (ZFC), and Field Cooled (FC) measurements in the presence of the magnetic field of 20Oe and results are found to be in order that should be observed for a superconductor. The Raman spectroscopy has been carried out using the Jobin-Yvon T64000 Triple-mate Raman spectrometer. It is equipped with a 514 nm laser and detector as a CCD (charge-coupled device) system. The accuracy of the CCD is 0.5 cm$^{-1}$. The well-characterized sample has used to investigate the kinetic decay profile from a few femtoseconds to nanoseconds with the help of TRUS in reflection geometry. The HELIOS femtosecond transient absorption spectrometer has been used from dynamic measurements[35,36]. The ZnTPP (soon released as BND$^{TM}$) reference material calibrates its spectra in NPL. The pump laser and probed laser has generated using the mode-locked Micra laser and amplifier by Coherent. The wavelength selection has been carried out with the help of TOPAS, an optical parametric amplifier by Light Conversion. The exciting laser is a 35 fs pulse laser with a repetition rate of 1 kHz.

**Result and Discussion**

The single crystal has been synthesized using an optimized heat treatment, as described in our previous report [29]. The shiny silver crystal was grown through the process, and the cleaved flake has been taken to confirm the presence of (00l) planes. It has a rhombohedral crystal structure with $D_{3d}^5$ (R$\bar{3}$m) space group. The reflections confirm the single-crystalline phase of the Nb-doped sample. Figure 1 shows the XRD of the flake, exhibiting (0,0,3n) reflections plane with n=1,2,3,4,5, and 6. The crystal has purely oriented in the c-axis with no discernible impurity phases[33]. Comparing the Nb-doped $Bi_2Se_3$ crystal XRD with the $Bi_2Se_3$ XRD from the previous paper shows that there is an addition of (003) reflection planes. It is because of the intercalation of the Niobium atoms in between the quintuple layers of $Bi_2Se_3$. DC magnetization measurements have been made by using Quantum Design PPMS under Zero Field Cooled (ZFC), and Field Cooled (FC) measurements in the presence of the magnetic field of 20Oe and results are found to be in order that should be observed for a superconductor. ZFC measurements are made by first cooling the sample down to 2K and then applying the magnetic field of 20Oe in warming, and in the FC measurements sample is cooled down to 2K in the presence of a magnetic field. Superconductivity in the grown crystal is evident from clearly visible diamagnetic transition at 2.5K in both FC and ZFC measurements, as shown in the inset of figure 1. In the ZFC signal, this response is saturated near 2.2K, with a transition width of 0.3K. The response represents the quality of the sample to be good.

The second inset of figure 1 shows the comparison between the Raman spectra of Nb-doped and $Bi_2Se_3$ single crystal. Both the spectra are normalized to associate the different shifts in exact order. The Nb sample shows all the three Raman vibrational modes, as shown in the pure crystal, but with a slight change in the frequency modes. The $A_{1g}^1$ and $A_{1g}^2$ modes in Nb-doped are shifted from 131.09 cm$^{-1}$ and 174.39 cm$^{-1}$ to 129.85cm$^{-1}$ and 173.14 cm$^{-1}$ respectively from undoped TI. While the $E_g^2$ modes changed from 71.35cm$^{-1}$ to 71.66 cm$^{-1}$. Overall, there is a shift of Raman modes from higher frequency to smaller frequency due to the presence of an extra Columbic layer between quintuple layers[29,33]. The repulsion changes the bond length and angle as doping distorted the structure. The interaction suppressed the phonon vibrations modes, also due to which there is a bathochromic shift in the doped sample. The rhombohedral structured compounds show a change in the phonon frequency when doped with atoms[37,38]. Thereby, it is confirmed that by changing the composition, there are changes in the Raman modes due to the deformation of quintuple layers and structural deformation. It is well known that TIs show phonon oscillations when irradiated by a high intense Laser beam[12,29,39,40]. In the above Raman measurement, it is clear that Nb-doped $Bi_2Se_3$ shows redshift when irradiated with visible radiation. Here, this fact creates a possibility that a similar kind of shift can be observed with radiation of other frequency regions such as terahertz frequency range. Also, terahertz frequency oscillations have direct accordance with Raman $A^1_{1g}$ mode [12,29,40,41]. It makes this important to study the phonon oscillations in this range. For this purpose, TRUS measurements have carried out, as discussed.

The TRUS measurements have been carried out on the Nb-doped flake and the pure TI sample of $Bi_2Se_3$. The samples have been pumped with the exciting energy of 1.9 eV, and its kinetic decay is considered at 1.38 eV. The dynamic decay profile is viewed from a few femtoseconds to 3 nanoseconds, as shown in figure 2. It shows the relaxation of the excited carriers from their higher states to lower ones. It can easily be divided into three physiognomies part. Figure 2 shows three distinct parts of the kinetic decay profile. One is the trapping and exponential decay part. It shows the excitation of charge carriers from its ground state to higher excited states and thereby, relaxation of the excited carriers back to the ground state with different energy states. The free charge carriers produce the second part, and it termed as band renormalization. It arises due to many-body effects, which ascends due to exchange-correlation corrections. The remaining one is the oscillations part of the decay profile that occurs due to the transfer of energy from its excited carriers to phonons. It transpires at a smaller time scale that lasts up to 100 ps through this relaxation. It causes two types of vibrations at different time scales. One is at a higher time scale from 10 to 100 ps, which is a slow oscillation, while another is from few femtoseconds to 10 ps. The subsequent one is fast oscillations[11,12,40,42,43]. The insets of figure 2 show the fast oscillations as well as the slow oscillations. Both the decay profile shows two characteristics, one in the acoustic phonon vibrations and another is optical phonon vibrations. It resembles the coherent acoustic (CAP) and optical phonons (COP) modes, as shown earlier in $Bi_2Se_3$[12,39–44]. Here, we are considering the COP as it shows a significant potential of terahertz generation. The inset of figure 2 shows COP from 1ps to 10 ps. The Dopant profile shows a smaller magnitude of differential reflection signal and fewer COP oscillations as compared with the pure TI, as shown in the inset of figure 2. The lower signal resembles the shorter relaxation of the charge carriers. The fast oscillations are in accordance with the redshift of the frequency of $A^1_{1g}$ mode, as shown in Raman spectra.

Further, the COP of Nb-doped has been analyzed in terms of terahertz frequency. The two different approaches have used to analyze it from a kinetic decay profile. These are FFT (fast Fourier transformation) and FFD (filtering high-frequency component followed by fitting data)[12]. Figure 3 shows a comparison among the FFT analysis of the Nb-doped and pure sample. The kinetic profile from 0 ps to 10 ps has been considered as it corresponds to the COP. The cropped data has been taken to transform the time domain oscillations to the frequency domain through origin 9.1 software. The FFT signal processing command has been used for Fourier transformation. The terahertz frequency that comes out through this analysis is the same as 0.11 THz in Nb-doped and pure sample, respectively, as shown in figure 3. Another analysis that helps to extract terahertz is FFD. In this process, the cropped oscillations in the kinetic profile have been extracted out from the relaxation of excited charge carriers. The signal processing in origin 9.1 software has been used in which the FFT filter has been used. The data has been sieved out through high pass Fourier filter at 1.82 THz. The information has then fitted using a sinusoidal damped equation. Figure 4 shows the fitted data of the Nb-doped sample with this equation. The kinetic filter data of the Nb-doped sample has been fitted, and the fitting parameters are shown in Table I. The terahertz frequency that comes out with this analysis is 1.99 THz. Comparing the terahertz generated by the pure $Bi_2Se_3$ sample, the Nb-doped sample has small terahertz frequency as the pure sample shows a 2.14 THz, as shown in the previous report[12].

The prior FFT analysis does not unerringly give the terahertz as it includes the relaxation dynamics of excited charge carriers. The more reliable analysis is FFD. It shows that Nb doping in the $Bi_2Se_3$ reduces the terahertz frequency from 2.42 to 1.99 THz. The reason behind this shortening is in accordance with the redshift of the $A_{1g}^1$ modes in Raman spectra. The Raman mode corresponds to the COP oscillations. The COP vibrations generated terahertz frequency. The shift is due to the presence of opposite signs in quintuple layers. It has been associated with shortening and stretching of bond chain and angle. The deformation of layers has been observed in Rietveld analysis. The PXRD of the crushed sample shows an increase in the C parameter, as shown in our previous report[33]. The Raman spectral shift and terahertz frequency shift are due to these structural deformations. This shift confirms the intercalation of Nb atoms. The expansion of C parameters reveals the stretching of the quintuple layers due to the dopant atoms. The chain expansion can be interpreted as the Nb atoms are intercalated between the Quintuple layers. These atoms form a layer that interacts with the quintuple layers, and the van der Waals force of attraction among layers cannot stop it. Thereby, stretching the chain length as Nb atoms has intercalated between every pair of quintuple layers. These Nb atoms slightly deform the structure. The 0.25 Nb intercalated in the pure $Bi_2Se_3$ TI, has caused a redshift in Raman modes and tuned the terahertz generation to a shorter frequency.

**Conclusion**

The report shows the formation of the $Nb_{0.25}Bi_2Se_3$ topological insulator. The single crystalline nature has been from the (00l) reflection planes. It has been concluded that with the Nb dopant, there is stretching of the quintuple layers. This distending deforms the structure as increasing the bond angle and bond length. There is also a redshift in the doped system, as shown by the Raman spectra. The modes redshift is due to the same deformation, as also shown in the Rietveld analysis. Superconductivity ingrown crystal is evident from clearly visible diamagnetic transition at 2.5K in both FC and ZFC measurements. In the ZFC signal, this response is saturated near to 2.2K with a transition width of 0.3K. It represents the quality of the sample to be good. The doped sample has further analyzed as a capable candidate for terahertz generation. The FFD analysis shows a 1.99 THz generation, which has tuned due to the presence of the dopant atom. The Nb atom intercalated between the quintuple layers and change the generating terahertz. Therefore, by doping with a suitable atom, we can tune the terahertz. Thus, $Bi_2Se_3$ can become a potential candidate for future application of generation of tunable terahertz frequency by dopant atoms.

**Acknowledgment**

The director of CSIR-NPL highly supports the project. The authors would like to thanks Dr. Nita Dilawar and Mr. Jasveer Singh for Raman spectroscopy and Mr.K.M Kandpal for vacuum sealing of sample in quartz tube for heat treatment. Mr. Prince Sharma likes to thanks CSIR-NPL and AcSIR for enrollment as Senior research scholar and CSIR-UGC for financial support.

**Figure Caption**

Figure 1. Single crystal PXRD of the flake of Nb0.25Bi2Se3 crystal and insets showing an FC and ZFC measurements and the Raman spectral comparison of Nb-doped sample and Pure sample.

Figure 2. The kinetic decay profile at 894 nm (1.38 eV), which has pumped at 1.9 eV. It covers an extensive decay from few femtoseconds to 3 ns. The primary inset shows the CAP as ranged from 1-100 ps, and the second shows the COP as from 1-10 ps.

Figure 3. FFT analysis of the Nb-doped and pure TI showing the frequency of terahertz generation.

Figure 4. Fitted FFT high pass filter data at 1.82 THz with the damped sinusoidal equation in FFD analysis.

Table I. Sinusoidal fitted parameters in FFD analysis.

**Figure 1.**

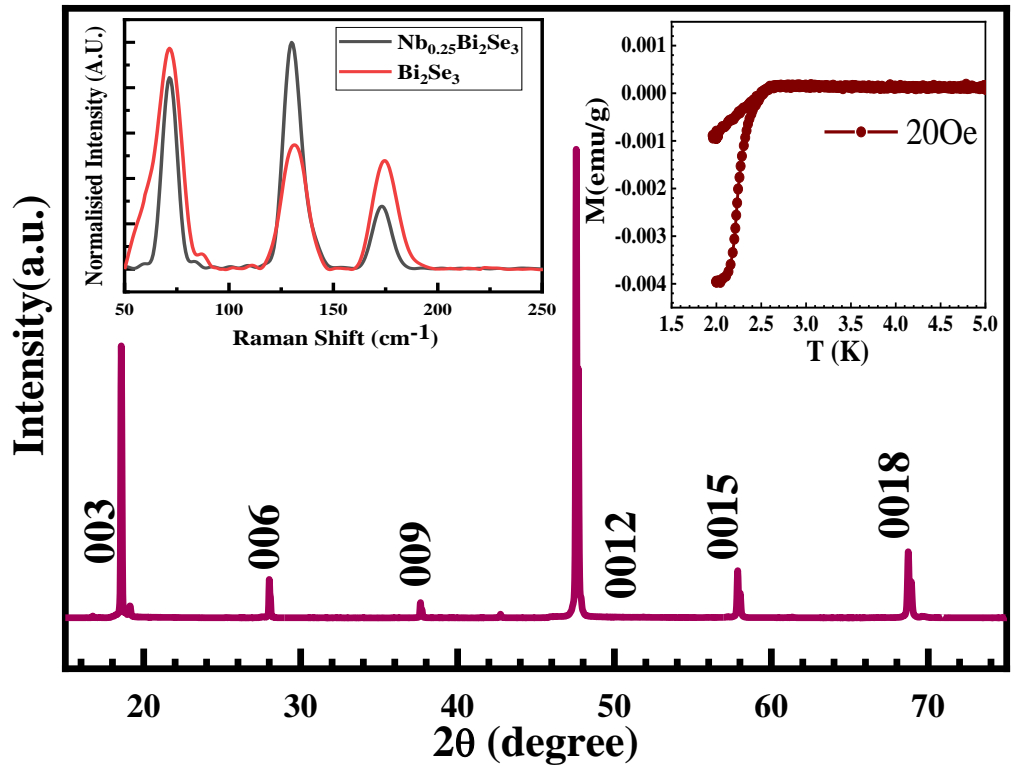

**Figure 2.**

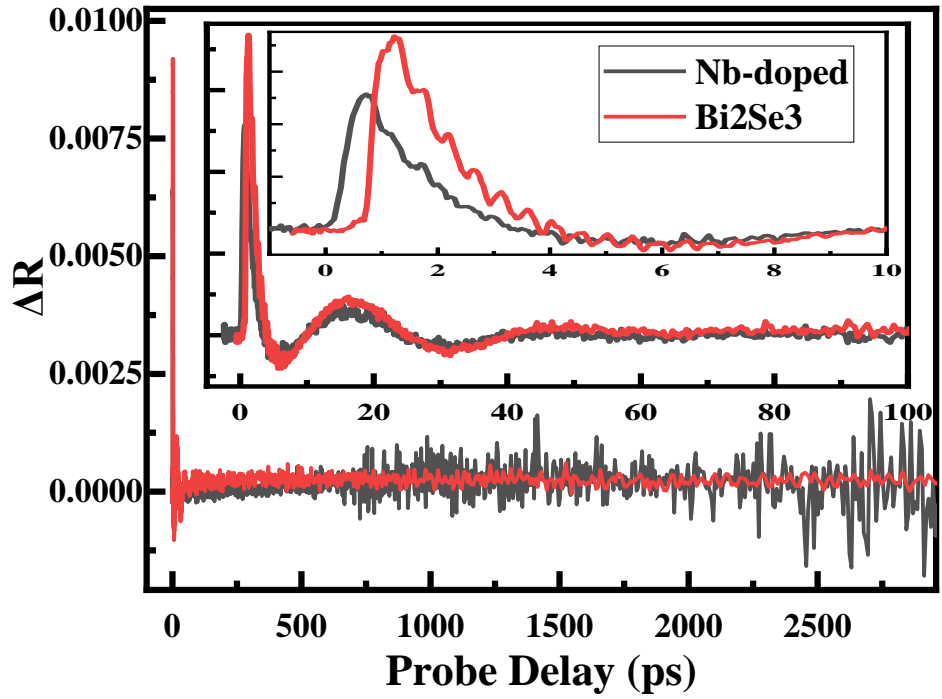

**Figure 3.**

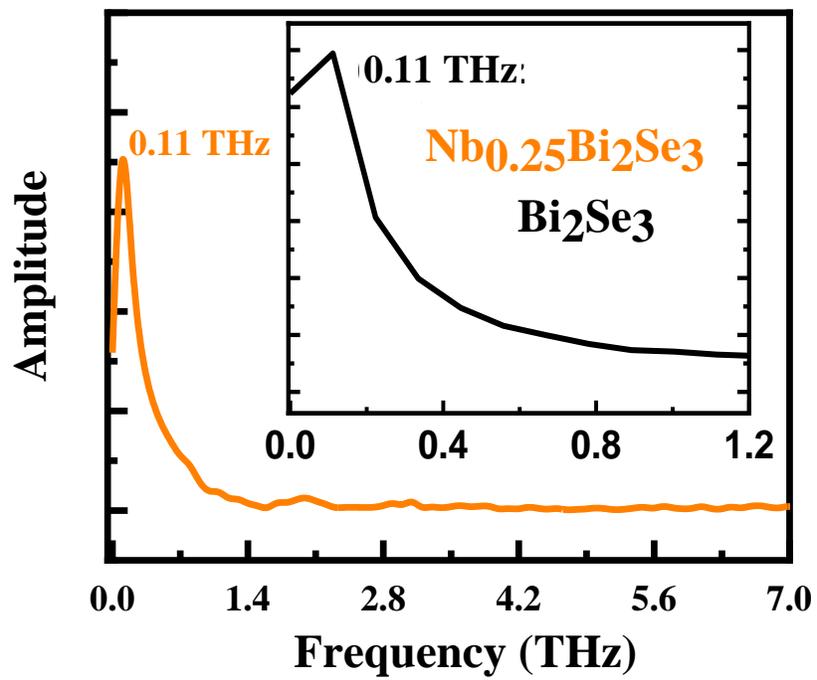

**Figure 4.**

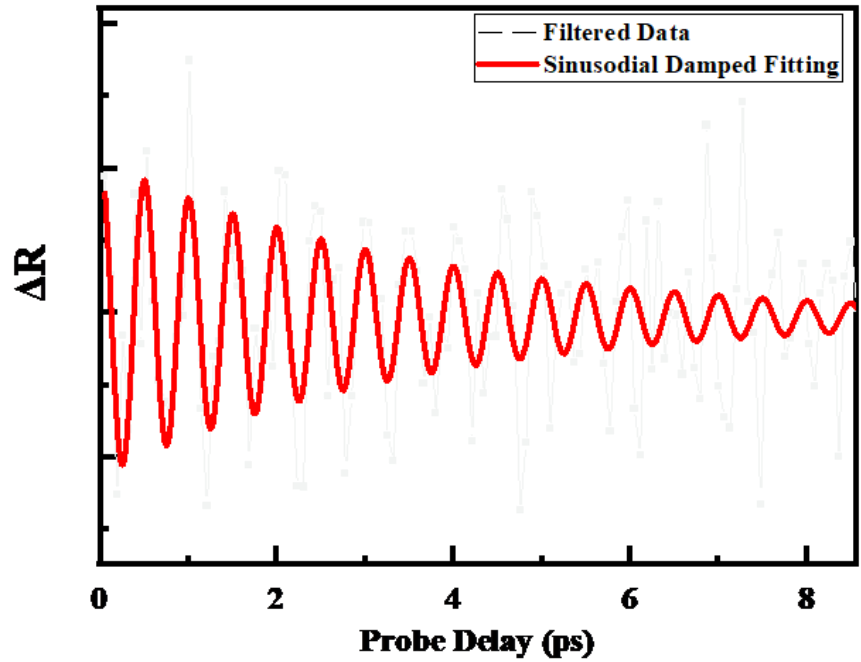

Table I.

| Model | SineDamp |
|---|---|
| Equation | y=y0 + A*exp(-x/t0)*sin(pi*(x-xc)/w) |
| Plot | Filtered Y1 |
| y0 | 7.92933E-4 ± 8.32461E-6 |
| xc | -0.11796 ± 0.0153 |
| w | 0.25005 ± 0.00152 |
| t0 | 3.55677 ± 0.97708 |
| A | 2.18939E-4 ± 4.14109E-5 |
| Reduced Chi-Sqr | 1.01664E-8 |
| R-Square (COD) | 0.29653 |
| Adj. R-Square | 0.27672 |